\documentclass[twocolumn]{webofc}

\usepackage[varg]{txfonts}   
\def\bra{\,<\!} \def\ket{\!>\,} \def\ack{\,|\,}
\begin{document}
\title{Microscopic aspects of $\gamma$ softness in atomic nuclei }
%
%

\author{\underline{\firstname{Stefan} \lastname{Frauendorf}}\inst{1}\fnsep\thanks{\email{sfrauend@nd.edu}} 
\and
        \firstname{Gowhar }\lastname{Bhat}\inst{2,3}
         \and
        \firstname{Nazira} \lastname{Nazir}\inst{4}
        \and
         \firstname{Niyaz} \lastname{Rather}\inst{5}
         \and
          \firstname{Syed} \lastname{Rouoof}\inst{5}
          \and
          \firstname{Sheikh } \lastname{Jehangir}\inst{5}
          \and
           \firstname{Javid} \lastname{Sheikh}\inst{4}
         .
}
\institute{
Department of Physics and Astronomy , University of Notre Dame, Notre Dame, Indiana 46556,  USA 
\and
Department of Physics, SP College  Srinagar, Jammu and Kashmir, 190 001, India
\and
Cluster University Srinagar, Jammu and Kashmir, Srinagar, Goji Bagh, 190 008, India
\and
Department of Physics, University of Kashmir, Srinagar, 190 006, India
\and
Department of  Physics, Islamic University of Science and Technology, Awantipora, 192 122, India
}
\abstract{%
It is demonstrated that the Triaxial Projected Shell Model reproduces the energies and transition probabilities of the
 nucleus $^{104}$Ru and the rigid triaxial nucleus $^{112}$Ru. An interpretation in terms  of band mixing is provided.
}
\maketitle
%
My  presentation covers  published work \cite{Jehangir,Ruoofa}, a paper under review \cite{Ruoofb} and ongoing work \cite{Ruoofc}.
I start with discussing the concept of $\gamma$ softness vs. $\gamma$ rigidness and the corresponding observables in the traditional frame work
of a phenomenological Bohr Hamiltonian. Then I sketch the  Triaxial Projected Shell Model (TPSM) and compare  calculations for $^{104,122}$Ru with experiment.
 Finally I discuss the TPSM   interpretation of the appearance $\gamma$ softness or $\gamma$ rigidness, which is an anternative perspective
 traditional concept of a collective Hamiltonian.


The notation of $\gamma$ softness came into use in the context of the Bohr Hamiltonian (BH) which describes collective states of  the quadrupole distortion of
the nuclear surface \cite{BMII}, where the parameter $\beta$ describes the deformation and $\gamma$ the triaxiality. The simple Gamma Rotor Hamiltonian
of Ref. \cite{caprio11} is one possibility to quantify the concept of $\gamma$ softness. The  deformation $\beta$ is assumed to be fixed. The scaled BH in the 
$\gamma$ degree of freedom
\begin{equation}\label{eqn-BH}
\hat\Lambda^2+V(\gamma)=\hat\Lambda^2-\chi \cos3\gamma+\xi\cos^23\gamma
\end{equation}
contains a potential determined by the parameters $\chi$ and $\xi$ and a kinetic term
\begin{equation}
\label{eqn-Lambdasqr}
\hat\Lambda^2=-\biggl(
\frac{1}{\sin 3\gamma} 
\frac{\partial}{\partial \gamma} \sin 3\gamma \frac{\partial}{\partial \gamma}
- \frac{1}{4}
\sum_{i=1,2,3} \frac{\hat{L}_i^{\prime2}}{\sin^2(\gamma -\frac{2}{3} \pi i)}
\biggr),
\end{equation}
which is the sum of the $\gamma$ kinetic energy and the rotational energy.

\begin{figure}
\centering
\includegraphics[width=\linewidth,clip]{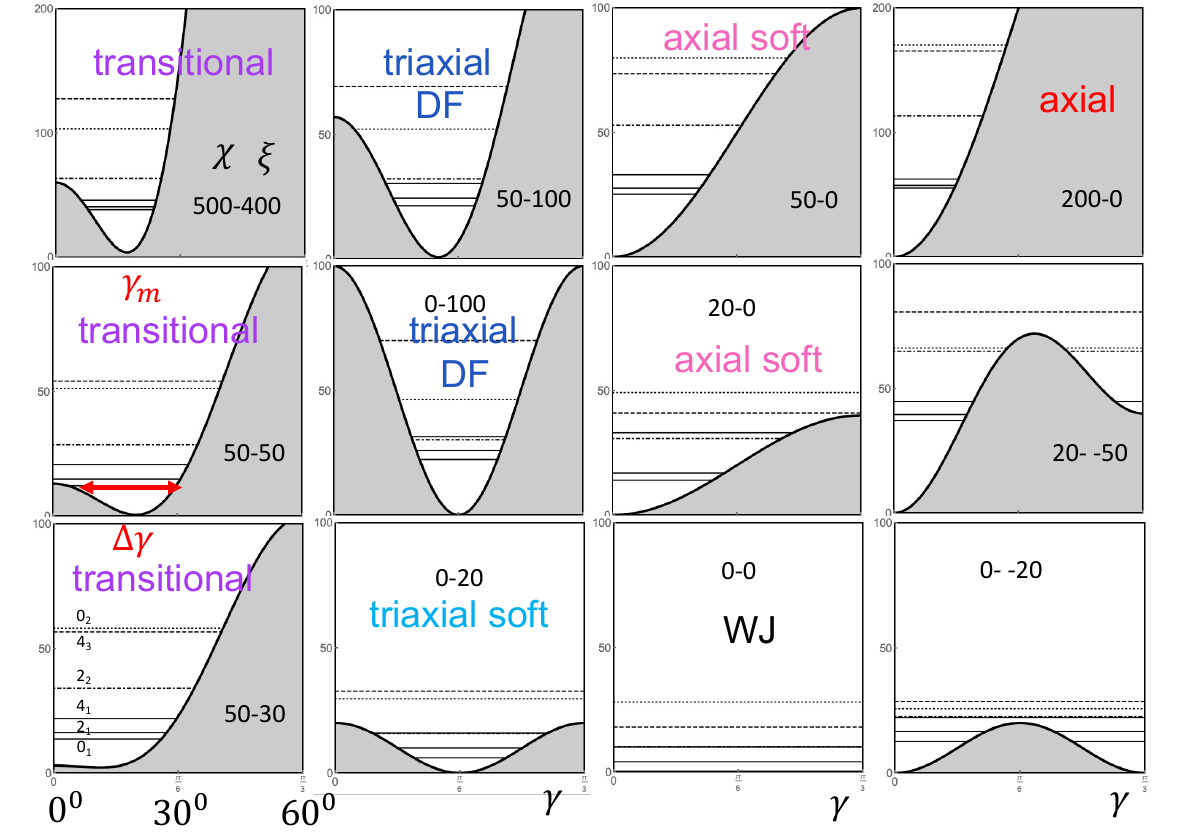}
\caption{Selection of potentials given by  Eqn. (\ref{eqn-BH}). The potential parameters are included as $\chi-\xi$.
The energies of the lowest states are shown as horizontal lines, where the line type indicates 
the $I_n$ of the state (see the panel 50-30). The potential minimum at $\gamma_m$ is a measure of the 
triaxiality and the width $\Delta \gamma$ is a measure how soft the potential is. To simplify terminology, "axial" or "triaxial" denote
the cases with a small width. DF stands for the approach to rigid triaxial rotor (Dawydov-Fillipov limit)  and W J
for the $\gamma$ independent case (Wilet-Jean). Modified from Ref. \cite{Ruoofb}.}
\label{fig-potentials}       
\end{figure}

\begin{table}[h]
\caption{Triaxiality characteristics of the potentials plotted in Fig. \ref{fig-potentials} .
The staggering parameter $\bar S(I)$ is defined by Eq. (\ref{eqn-bS(I)}). The excitation energy $E(2^+_1)_{GR}$  is
 the difference $E(2^+_1)-E(0^+_1)$ of the eigenvalues of the Gamma Rotor Hamiltonian (\ref{eqn-BH}).
}
\resizebox{\linewidth}{!}
  {
\begin{tabular}{|c|c|c|c|c|c|c|c|c|}
  \hline
 $\chi-\kappa$  &$E(2^+_1)_{GR}$ &$\gamma_m$	 &$\Delta \gamma$	 &${\frac{E(2^+_2)}{E(2^+_1)}}$	&${\frac{E(2^+_2)}{E(4^+_1)}}$	
 &$\bar S(6)$ &$Q(2^+_1)$&$\begin{array}{l} B(E2,2^+_2 \\ \rightarrow 0^+_1 )\end{array}$\\
   \hline		
100-0        &2.24       & 0       &17               &18.0    &3.31    &-0.14 	  	 &-0.878        &0.047	\\    %
50-0          &2.39      &0          &20                &11.6   &3.55     &-0.51               &-0.861     &0.064\\ %
20-0         &2.87        &  0   & 24                &5.82      &1.93      &-1.75 		 &-0.797       &0.079 \\  %
0-0           &4.00         &30     &60              &2.50     &1.00         &-2.75 	&0.000         &0.000  \\  %
0-200        &3.05       & 30       &16               &2.11     &0.81     &3.87		 &0.000        &0.000     \\   %
0-100       &3.57      &30        &19               &2.21     &0.86      &3.12  		 &0.000        &0.000   \\ %
0-20          &3.95      & 30    &22                  &2.45    &0.98       &0.50 		 &0.000        &0.000  \\ %
50-100     &3.10       &25          &19	 	   &3.50   &1.20     &1.85 		 &-0.693      &0.084  \\%
500-400	&2.34	&17	        &15		&10.8   &3.17          &0.32  	&-0.861      &0.066	\\%
50-30        & 2.58     &11     &26                  &7.78    &2.44       & -0.35  	&-0.835       &0.082  \\%
0- -20       &3.39      &30     &35                 &2.42     &0.96        &-5.79		&0.000        &0.000\\%
20- -50      &2.35     &34     &17                   &11.74     &3.62    &-5.55    	&-0.868      &0.032  \\%
\hline								
\end{tabular}
}
\label{tab-gav}
\end{table}

Fig. \ref{fig-potentials} shows examples of various potentials, which illustrate the notations of
rigid vs. soft and triaxial vs. axial. Tab. \ref{tab-gav}  lists the main characteristics of the collective mode.
 The position of the minimum (maximum) of the potentials $\gamma_m$ quantifies the degree of triaxiality.
The softness of the potentials $\Delta \gamma$ is quantified as the length of the bar E($0^+_1)<V(\gamma)$, 
which is a measure  of the ground-state fluctuation in $\gamma$. 
The next  twocolumns
list important energy criteria, which characterize the nature of triaxiality, 
namely the ratios ${\left[\frac{E(2^+_2)}{E(2^+_1)}\right]}$, ${\left[\frac{E(2^+_2)}{E(4^+_1)}\right]}$.

The staggering parameter	 $S(I)$ of the $\gamma$ band is an essential observable indicating the $\gamma$ softness \cite{NV91}.
 \begin{eqnarray}
S(I)= [E(I)-2E(I-1)]+E(I-2)]/E(2^{+}_1),\label{eqn-S(I)}\\
\bar S(I)=[S(I)-S(I+1)]/2.\label{eqn-bS(I)}
\end{eqnarray}
The odd-$I$-down pattern indicates the concentration of the collective 
wave function around a finite $\gamma$-value (static triaxiality), whereas the even-I-down pattern points to a spread of the wave function over a large range 
of $\gamma$ (dynamic triaxiality).
Tab. \ref{tab-gav} lists the modified staggering parameter $\bar S(6)$.

\begin{figure}[t]
\centering
\includegraphics[width=\linewidth,clip]{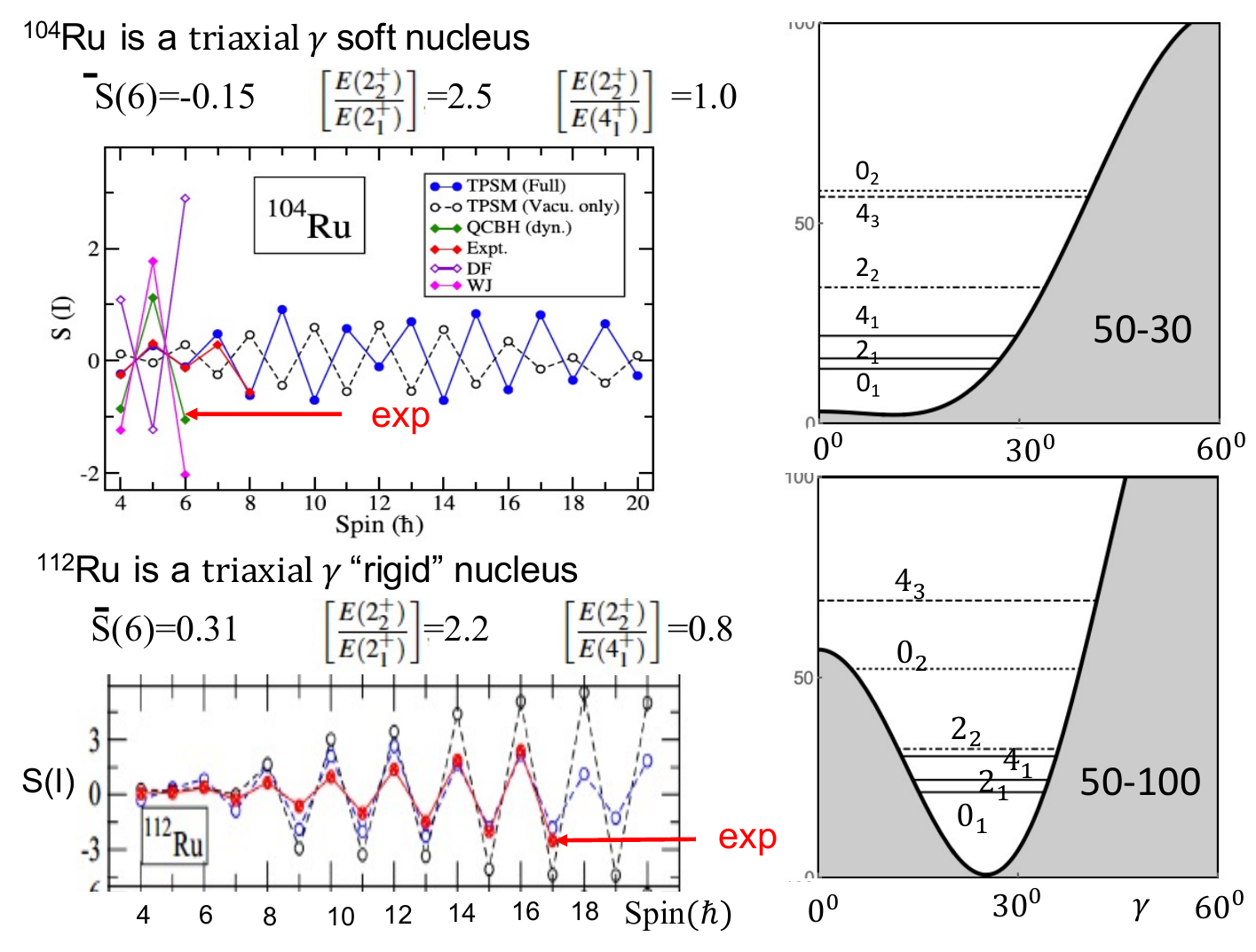}
\caption{Phenomenlogical potentials  (see Fig. \ref{fig-potentials}) which best reproduce the shown
experimental signatures of triaxiality in $^{104,112}$Ru (see Tab. \ref{tab-gav}). 
Combination from Refs. \cite{Jehangir,Ruoofa,Ruoofb,Ruoofc}.
}
\label{fig-phenRu}       
\end{figure}

Fig. \ref{fig-phenRu} shows the experimental signatures of triaxiality in $^{104,112}$Ru. Added are the two potentials from Fig. \ref{fig-potentials}, which
according to Tab. \ref{tab-gav} provide the best description of the energetic characteristics. Clearly, $^{104}$Ru is a very $\gamma$ soft nucleus while
the triaxial shape is well established in $^{112}$Ru.

The details of the TPSM are described in Refs. \cite{Jehangir,Ruoofa,Ruoofb,Ruoofc}. 
The basic methodology of the TPSM approach is similar to that of the standard spherical
shell model with the exception that angular-momentum projected deformed basis
is employed to diagonalize the shell model Hamiltonian.
The Hamiltonian consists of monopole pairing, quadrupole pairing, and
quadrupole-quadrupole interaction terms within the configuration space of three major oscillator shells
($N=3,4,5$ for
neutrons and $N=2,3,4$ for protons) 
\begin{equation}
\hat H = \hat H_0 - {1 \over 2} \chi \sum_\mu \hat Q^\dagger_\mu
\hat Q^{}_\mu - G_M \hat P^\dagger \hat P - G_Q \sum_\mu \hat
P^\dagger_\mu\hat P^{}_\mu ,
\label{hamham}
\end{equation}
where $\hat H_0$  is the spherical single-particle potential \cite{BMII}.
The pairing constant  $G_M$  is determined by the BSC relation \mbox{$G_M\bra\hat P\ket=\Delta$}. 
The quadrupole pairing strength
$G_Q$ is assumed to be 0.18 times $G_M$. The QQ-force
strength $\chi$ is obtained by the self-consistency 
relation \mbox{$2 \hbar\omega\varepsilon= 3\chi \bra\Phi\ack\hat Q_0\ack\Phi\ket$}.
The shell model Hamiltonian (\ref{hamham}) is diagonalized in the space of angular-momentum projected
multi-quasiparticle configurations generated by the triaxial version of the Nilsson Hamiltonian \cite{BMII} with
the deformation parameters $\varepsilon$ and $\gamma$.
The basis is composed of the 0-qp vacuum, the two-quasiproton, the two-quasineutron and the combined
four-quasiparticle configurations
\begin{eqnarray}\label{basis}
\hat P^I_{MK}\ack\Phi\ket,~
\hat P^I_{MK}~a^\dagger_{p_1} a^\dagger_{p_2} \ack\Phi\ket,~
\hat P^I_{MK}~a^\dagger_{n_1} a^\dagger_{n_2} \ack\Phi\ket,\nonumber\\
\hat P^I_{MK}~a^\dagger_{p_1} a^\dagger_{p_2}
a^\dagger_{n_1} a^\dagger_{n_2} \ack\Phi\ket,
\end{eqnarray}
The TPSM has $\Delta_{p,n},~\varepsilon$ and $\gamma$ as input parameters. 
The pairing gaps  $\Delta_{p,n}$ are chosen such that the overall observed odd-even mass differences are
reproduced for nuclei in the considered  region. The deformation parameter 
$\varepsilon$ is taken from $B(E2, 2_1^+\rightarrow 0_1^+ )$ systematics or mean field equilibrium deformations.
The triaxiality parameter $\gamma$ is adjusted to reproduce the $\gamma$ band head energy $E(2^+_2)$.

\begin{figure}[h!]
\centering
\includegraphics[width=1\linewidth,clip]{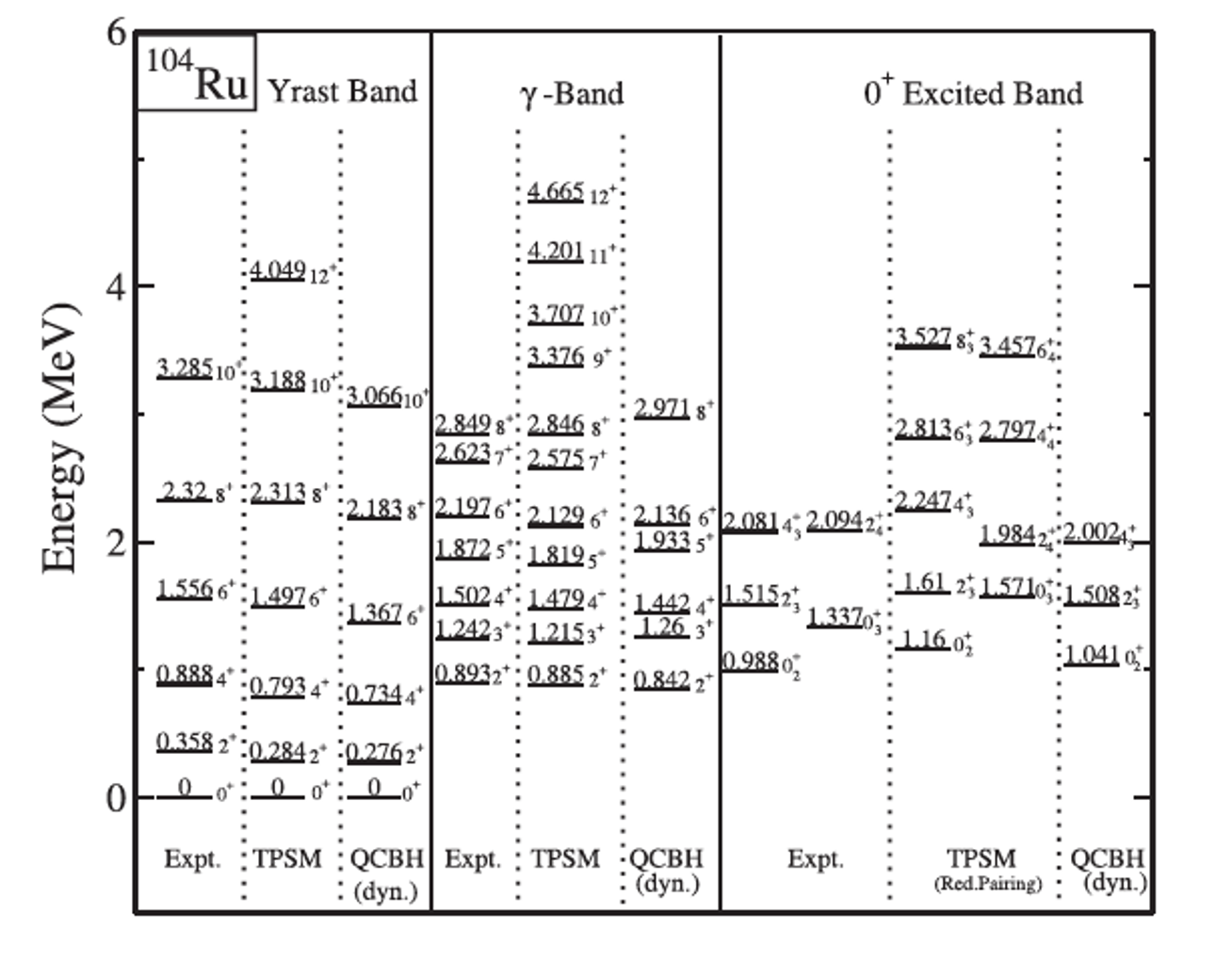}
\caption{Comparison of the experimental levels  of $^{104}$Ru with the calculations 
by means of the TPSM and the microscopic Bohr Hamiltonian   \cite{Ru104} (QCBH dyn). From Ref. \cite{Ruoofa}.
}
\label{fig-levels}       
\end{figure}

Fig. \ref{fig-levels} demonstrates that the TPSM very well reproduces the experimental energies of the ground, $\gamma$ and
excited $0^+$ bands. In particular, the even-I-down pattern of the staggering parameter $S(I)$ of the $\gamma$ band
is quantitatively reproduced, which is the energy signature $\gamma$ softness (see Fig. \ref{fig-phenRu}).  
 Tables \ref{tab-1}, \ref{tab-2} compare the TPSM with the large number of reduced $E2$ and $M$
 matrix element measured in the COULEX experiment of Ref. \cite{Ru104}.
 The agreement of the TPSM results with the COULEX data is remarkable, 
 because the $E2$ matrix elements provide the most direct information on the statics and dynamics of the collective quadrupole
 modes.  Comparing the "Full" matrix elements with "Vacu."  ones shows that the quasiparticle admixtures generate 
 substantial shifts that generate the good agreement with the experiment.  

\begin{figure}
\centering
\includegraphics[width=0.7\linewidth,clip]{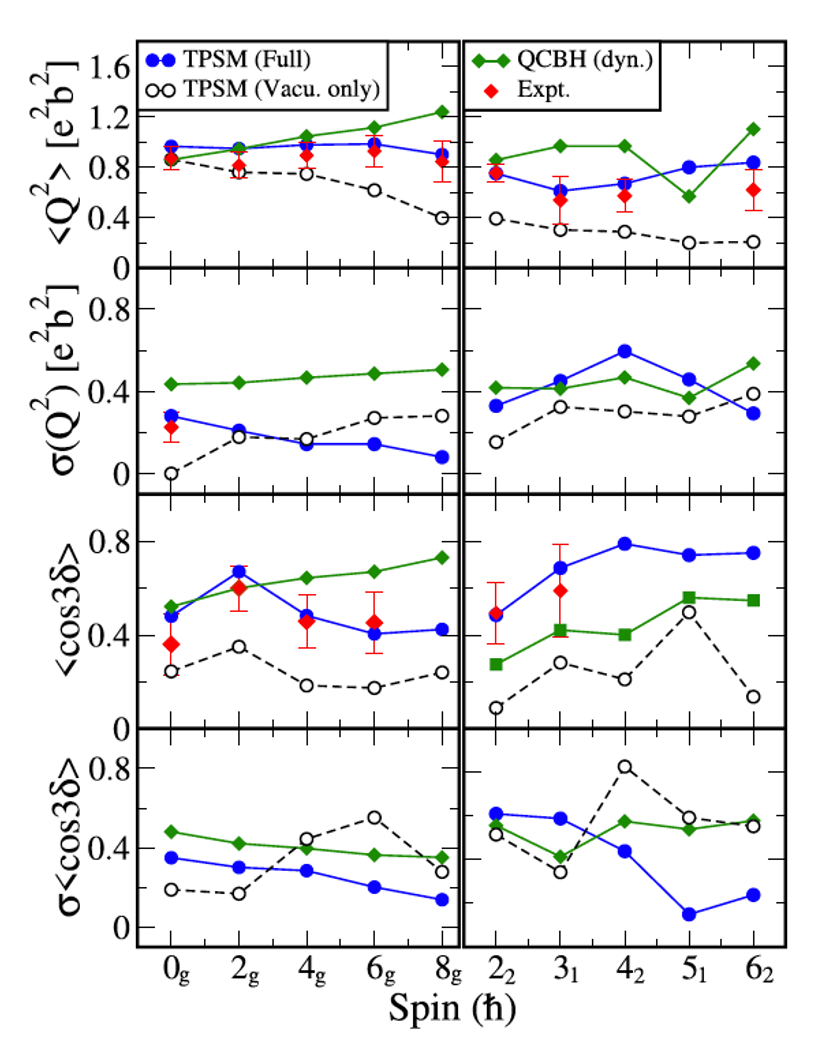}
\caption{Comparison of the observed (Expt.) with the TPSM  shape invariants
for the ground-band (left panels) and $\gamma$-band (right panels) in $^{104}$Ru.
The results from the microscopic Bohr Hamiltonian \cite{Ru104}  (QCBH dyn) are included.   From      Ref. \cite{Ruoofa}. }
\label{fig-invRu104}       
\end{figure}

\begin{figure}
\centering
\includegraphics[width=0.7\linewidth,clip]{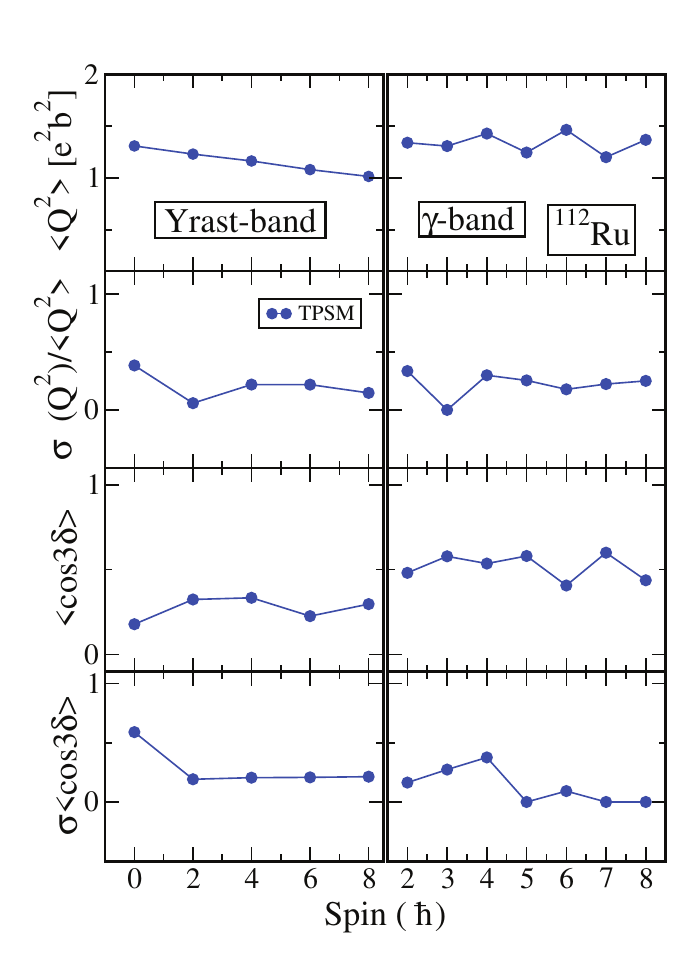}
\caption{As Fig. \ref{fig-invRu104} for $^{112}$Ru. From Ref.\cite{Ruoofc}.}
\label{fig-invRu112}       
\end{figure}

Figs. \ref{fig-invRu104} and  \ref{fig-invRu112} depict the quadrupole shape invariants  \cite{DC86} and their dispersions calculated from the reduced $E2$ matrix elements. 
 The invariant $\bra I^+_n | Q^2 | I^+_n \ket$ measures the average intrinsic deformation of a state $I^+_n$. 
 The invariant $\bra I^+_n |\cos 3 \delta | I^+_n \ket$ contains the information about the triaxiality of the intrinsic shape, 
 where $\delta=\arctan{\bra I^+_n |\cos 3 \delta | I^+_n \ket}$. The invariant $Q$  is  approximately  
 proportional to $\beta$,  and the invariant $\delta$ is approximately equal to
   $\gamma$ of the liquid drop model \cite{BMII}. 
   
  The ground-band TPSM values of $\bra \cos 3 \delta\ket \sim 0.5$ in  Fig. \ref{fig-invRu104} signify 
   a substantial triaxiality with preference for
 prolate shape, corresponding to $\gamma\approx20^\circ$. 
 The TPSM dispersion $\sigma<\cos 3 \delta> \sim 0.3$  indicates large 
  fluctuations of the triaxiality parameter with 70\% of the distribution within the range $9^\circ < \gamma<24^\circ$. Hence, the TSPM shape invariants
  imply that $^{104}$Ru is $\gamma$ soft, which is in accordance with the experimental staggering parameter $S(I)$  and $\bar S(6)=-0.15$ in Fig. \ref{fig-phenRu}.
 
  The shape invariants for $^{112}$Ru in Fig. \ref{fig-invRu112} correspond to  narrow distributions of  $7^\circ$ width centered at $\gamma\approx 26^\circ$ for the ground band and
  $\gamma\approx 20^\circ$ for the $\gamma$ band. They indicate a well established triaxial shape, which is consistent with the experimental 
  even -I- up pattern of $S(I)$ and $\bar S(6)=0.31$ in Fig. \ref{fig-phenRu}.

\begin{table}
\caption{Comparison of all known experimental reduced  $E2$ diagonal, in-band and inter-band matrix elements 
 $\bra I_i || E2 || I_f\ket(e.b)$, (associated errors in parenthesis) and calculated ones for yrast- and $\gamma$-band of
    $^{104}$Ru. From Ref. \cite{Ruoofa}.
    }
  
\resizebox{\linewidth}{!}
  {
\begin{tabular}{|c|c|c|c||c|c|c|c|}
  \hline
  $I_i \rightarrow I_f$   & Expt.    &TPSM       &TPSM
  & $I_i \rightarrow I_f$ & Expt.    &TPSM       &TPSM
  \\
                          &         &(Full)     &(Vacu.)
  &                       &         &(Full)     &(Vacu.)                
  \\

  \hline
  $2_1\rightarrow 2_1$    &-0.71(\textit{11)}             &-0.817  &-0.634
  &$4_2\rightarrow 3_1$   &$\pm $0.68  (\textit{5)}       &-0.787  &-0.597 \\
  $4_1\rightarrow 4_1$    &-0.79(\textit{15)}             &-0.906  &-0.437
  &$5_1\rightarrow 3_1$   &1.22(\textit{4)}               &1.184   &0.697\\
  $6_1\rightarrow 6_1$    &-0.70(\textit{$^{+30}_{-20}$ })  &-0.868  &-0.342
  &$6_2\rightarrow 4_2$   &1.52 (\textit{12)}             &1.521   &0.682\\ 
  $8_1\rightarrow 8_1$    &-0.6(\textit{$^{+3}_{-5}$ })     &-0.855  &-0.297
  &$8_2\rightarrow 6_2$   &2.02(\textit{4)}               &2.056   &0.747\\
  $2_2\rightarrow 2_2$    &0.62(\textit{8)}               &0.648   &0.633
  &$2_2\rightarrow 0_1$   &-0.156 (\textit{2) }           &-0.141  &-0.225\\
  $4_2\rightarrow 4_2$    &-0.58(\textit{18)}             &-0.749  &-0.534
  &$2_2\rightarrow 2_1$   &-0.75(\textit{4) }             &-0.722  &-0.612\\
  $6_2\rightarrow 6_2$    &$\pm$1.0(\textit{3)}           &-1.105  &-0.763
  &$2_2\rightarrow 4_1$   &$\epsilon$ [-0.1, 0.1]         &-0.090  &-0.001\\
  $2_1\rightarrow 0_1$    &0.917 (\textit{25)}            &0.973   &0.901
  &$3_1\rightarrow 2_1$   &0.22(\textit{10)}              &0.254   &0.302\\ 
  $4_1\rightarrow 2_1$    &1.43 (\textit{4)}              &1.591   &1.456
  &$3_1\rightarrow 4_1$   &-0.57                          &-0.517  &-0.559\\
  $6_1\rightarrow 4_1$    &2.04 (\textit{8)}              &2.081   &1.830  
  &$4_2\rightarrow 2_1$   &-0.107 (\textit{8 ) }          &-0.113  &-0.054 \\  
  $8_1\rightarrow 6_1$    &2.59 (\textit{$^{+24}_{-9}$ )}   &2.486   &1.902
  &$4_2\rightarrow 4_1$   &-0.83(\textit{5) }             &-0.840  &-0.505 \\
  $10_1\rightarrow 8_1$   &2.7 (\textit{6)}               &2.668   &1.623
  &$6_2\rightarrow 4_1$   &-0.22(\textit{$^{+6}_{-12}$ })   &-0.230  &-0.682\\
  $3_1\rightarrow 2_2$    &-1.22 (\textit{10)}            &-1.241  &-0.935
  &$6_2\rightarrow 6_1$   &$>$-0.84                       &-0.947  &-0.411 \\
  $4_2\rightarrow 2_2$    &1.12(\textit{5)}               &1.095   &0.510
  &                      &                               &        &      \\

  \hline

\end{tabular}
}
 \label{tab-1} 
\end{table}
\begin{table}
\caption{Comparison of all known experimental reduced  $E2$ matrix elements 
 $\bra I_i || E2 || I_f\ket(e.b)$, diagonal, in-band  and inter-band  values (associated errors in parenthesis) and calculated ones for excited $0^+$ bands of
    $^{104}$Ru. From Ref. \cite{Ruoofa}.
    }
    \label{tab-2} 
\resizebox{\linewidth}{!}
  {
\begin{tabular}{|c|c|c||c|c|c|}
  \hline
$I_i \rightarrow I_f$ & Expt.    & TPSM       &$I_i \rightarrow I_f$ & Expt.  & TPSM   \\
  \hline
  $2_3\rightarrow 0_2$    &0.71(\textit{4)}                 &0.682   
  &$2_3\rightarrow 4_1$   & -0.370(\textit{4)}              &-0.311\\
  $4_3\rightarrow 2_3$    &0.75(\textit{25)}                &0.613  
  &$2_3\rightarrow 2_2$   &$\pm$0.22(\textit{$^{+25}_{-5}$})  &-0.237\\  
  $0_2\rightarrow 2_1$    &-0.266(\textit{8)}               & -0.221
  &$2_3\rightarrow 4_2$   &0.31(\textit{$^{+13}_{-6}$})       & 0.221\\ 
  $0_2\rightarrow 2_2$    &0.08 (\textit{3)}                & 0.099
  &$2_3\rightarrow 4_4$   &0.53(\textit{$^{+32}_{-14}$})      &0.481  \\ 
  $2_3\rightarrow 0_1$    &-0.071(\textit{3)}               &-0.048  
  &$0_3\rightarrow 2_1$   & $>$-0.1                         &-0.201 \\ 
  $2_3\rightarrow 2_1$    & $\pm$0.07(\textit{3)}           &-0.031
  &$2_3\rightarrow 2_3$    &-0.08(\textit{$^{11}_{25}$)}      &-0.631\\  

\hline
\end{tabular}
}
\end{table}

\begin{figure}
\centering
\includegraphics[width=0.9\linewidth,clip]{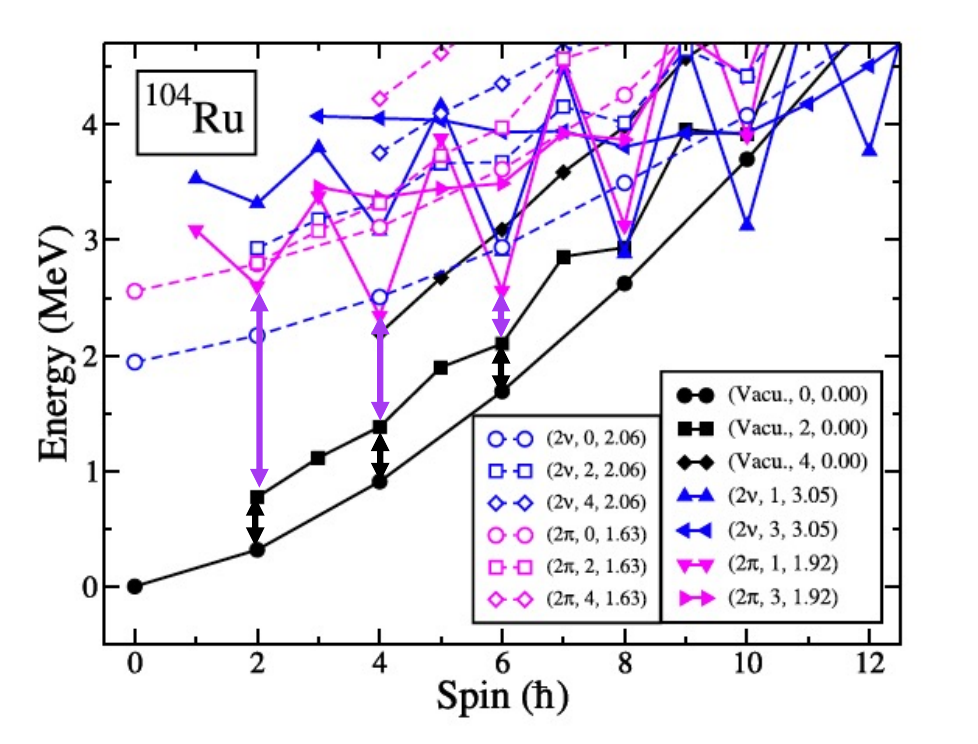}
\caption{TPSM projected energies before band mixing. The bands are labelled by
three quantities : quasiparticle character, $K$-quantum number and energy of the two-quasiparticle
state. For instance, $(2\pi, ~1, ~1.92)$ designates the $K=1$ state projected from
the $h_{11/2}$ two-quasiproton configuration with the energy of 1.92 MeV.
 The  $K=0, ~2,~4$ states projected from the quasiparticle vacuum are labelled with Vacu.
 The four-quasiparticle states lie above 5 MeV. The black arrows indicate the repulsion by the coupling to the ground band and 
 the purple arrows the repulsion by coupling to the even states of the $(2\pi, ~1, ~1.92)$ band after band mixing. Modified from Ref. \cite{Ruoofa}. }
  \label{fig-BandRu104}
       \end{figure}

\begin{figure}
\centering
\includegraphics[width=0.8\linewidth,clip]{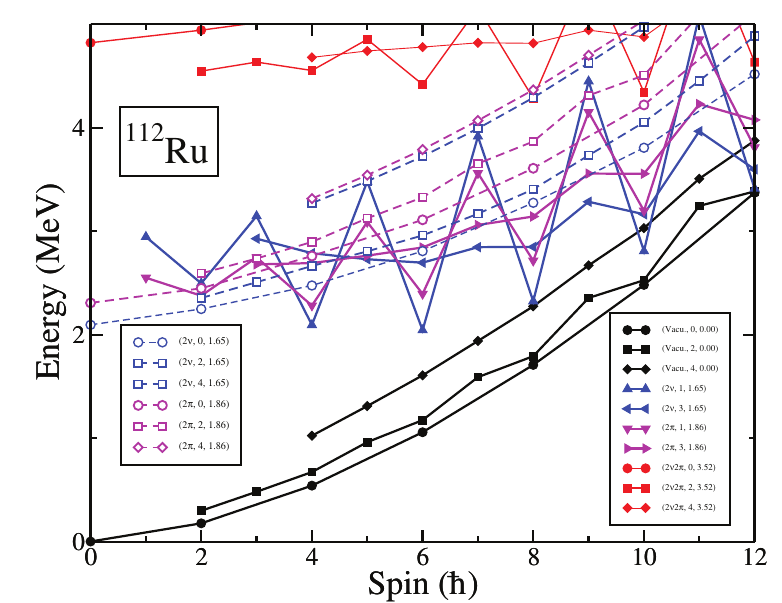}
\caption{As Fig. \ref{fig-BandRu104} for $^{112}$Ru. From Ref. \cite{Ruoofc}.}
 \label{fig-BandRu112}      
\end{figure}

The TPSM provides an alternative perspective on the experimental signatures of $\gamma$  softness.
The staggering parameter of the $\gamma$ band  is an important signature.
Fig. \ref{fig-BandRu104} displays the  energies of the projected quasiparticle configurations in $^{104}$Ru. 
Diagonalizing the Hamitonian within this basis generates repulsions between 
neighboring states. 

 The $K=2$ vacuum sequence, associated with the $\gamma$ band, shows the even-I-down pattern. 
Mixing it with the $K=0$ vacuum states, associated with the ground band, pushes  the even -I states of the 
$K=2$ band up. There is no shift for  odd I,  because the ground band contains only even-I states (see black arrows in Fig. \ref{fig-BandRu104}). 
The triaxial rotor  even-I-up pattern emerges for sufficiently 
strong interaction. This is the TPSM result when the basis is truncated to the 
$K$ states projected from zero quasiparticle state, which is shown as  "Vacu. only" (black dashed)
in Fig. \ref{fig-phenRu}. The same even-I-up are results found   for "Vacu. only" for all 35 studied nuclei \cite{Jehangir,Ruoofa}.

Admixing the lowest projected two quasiparticle states $(2\pi, ~1, ~1.92)$ (see purple arrows in Fig. \ref{fig-BandRu104})  and $(2\nu, ~1, ~3.05)$
pushes the even-I states of the  $K=2$  vacuum sequence down. There is less repulsion for odd I because of the larger distance.
This repulsion  prevails over the one by the ground band,  and the triaxial rotor pattern changes to the $\gamma$ soft pattern  even-I-down. The pattern flip is seen
in Fig. \ref{fig-phenRu} as the opposite staggering phases of the "Full" results (blue dashed) compared to the "Vacu. only" (black dashed) ones. 

Fig. \ref{fig-BandRu112} displays the  energies of the projected quasiparticle configurations in $^{112}$Ru.
The distance between the $K=2$ and $K=0$ sequences projected from the vacuum is much smaller than for $^{104}$Ru.
Their mixing is stronger and the corresponding shift is larger. It prevails over the downshift by the repulsion of the two-qusiparticle 
sequences, such the staggering amplitude is reduced but its phase remains unchanged (see Fig. \ref{fig-phenRu}).

The competition  between the two kinds of band mixing accounts for the observed $S(I)$ pattern for the 35 nuclides 
studied in Refs. \cite{Jehangir,Ruoofa,Ruoofb,Ruoofc}. The mixing with the two-quasiparticle states 
provides a natural explanation for the observed rapid change of the $S(I)$ with $N$ and $Z$, because 
the quasiparticle structure near the Fermi level plays a central role.
The mixing interpretation explains also  the $Z-N$ dependence of the triaxiality characteristics of transition probabilities \cite{Ruoofb}.

\begin{figure}[h!]
\centering
\includegraphics[width=0.88\linewidth,clip]{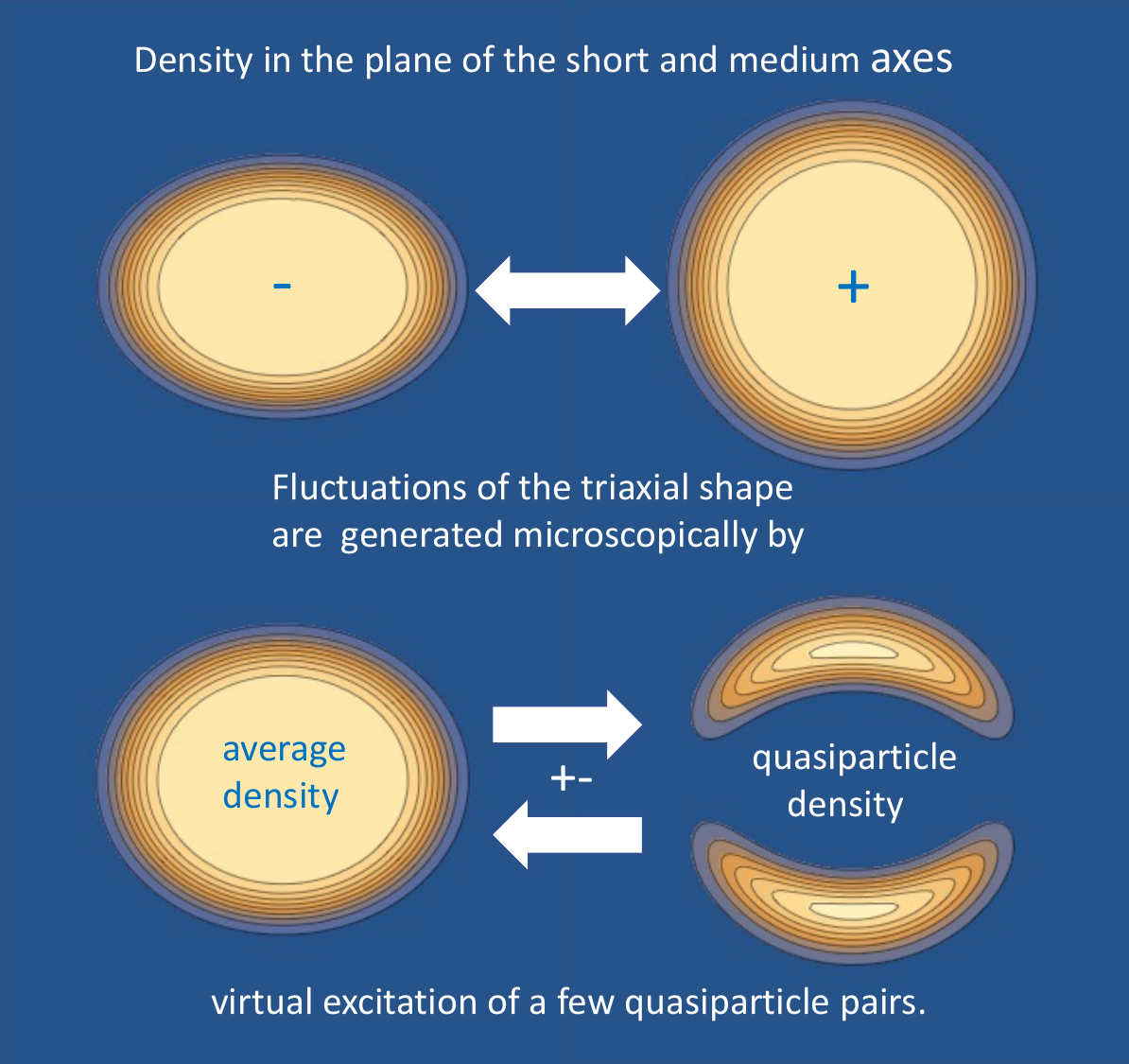}
\caption{Summary: One of the main research themes in quantum many-body systems is the emergence of collective features from 
microscopic degrees of freedom. Using the microscopic approach of the triaxial projected shell model, 
the authors demonstrate that admixing a few quasiparticle excitations into the vacuum configuration with a fixed triaxiality parameter 
$\gamma$ provides a quantitative description of the shape fluctuations of the 
$\gamma$-soft nucleus 
$^{104}$Ru. The collective features are elucidated using the quadrupole shape invariant analysis, and also 
the staggering phase classification of the $\gamma$ band. Figure and text from the editor's suggestion in Phys. Rev. C \textit{107}, Issue 2, (2023).}
\label{fig-2}       
\end{figure}

\end{document}